\documentstyle[12pt]{article}
\input psfig
\def\fun#1#2{\lower3.6pt\vbox{\baselineskip0pt\lineskip.9pt
        \ialign{$\mathsurround=0pt#1\hfill##\hfil$\crcr#2\crcr\sim\crcr}}}
\begin{document}
\thispagestyle{empty}
\renewcommand{\thefootnote}{\fnsymbol{footnote}}

\vspace*{-72pt}
\begin{flushright}
{\footnotesize
FERMILAB--PUB--97/069--A\\
OSU-TA-97-8 \\
astro-ph/9703119\\
\today}
\end{flushright}

\begin{center}
{\Large\bf Evolution of the Order Parameter \\ \vspace*{6pt}
after Bubble Collisions }

\baselineskip=14pt
\vspace{0.5cm}
Edward W.\ Kolb\footnote{Electronic address: {\tt  
rocky@rigoletto.fnal.gov}}\\
{\em NASA/Fermilab Astrophysics Center\\
Fermi National Accelerator Laboratory, Batavia, Illinois~~60510,  
and\\
Department of Astronomy and Astrophysics, Enrico Fermi Institute\\
The University of Chicago, Chicago, Illinois~~ 60637}\\
\vspace{0.4cm}
Antonio Riotto\footnote{Electronic address:
{\tt riotto@fnas01.fnal.gov}}\\
{\em NASA/Fermilab Astrophysics Center\\
Fermi National Accelerator Laboratory, Batavia, Illinois~~60510}\\
\vspace{0.1cm}
and\\
\vspace{0.1cm}
Igor I. Tkachev\footnote{Electronic address: 
{\tt tkachev@wollaston.mps.ohio-state.edu}}\\
{\em Department of Physics, The Ohio State University, Columbus, OH 43210, 
and \\ 
Institute for Nuclear Research of the Academy of Sciences of Russia
\\Moscow 117312, Russia}
\end{center}

\baselineskip=24pt

\begin{quote}
If a first-order phase transition is terminated by collisions of
new-phase bubbles, there will exist a period of nonequilibrium between
the time bubbles collide and the time thermal equilibrium is
established.  We study the behavior of the order parameter during this
phase.  We find that large nonthermal fluctuations at this stage tend
to restore symmetry, i.e., the order parameter is smaller than its
eventual thermal equilibrium value.  We comment on possible
consequences for electroweak baryogenesis.
\vspace*{8pt}

PACS number(s): 98.80.Cq, 11.27.+d
\end{quote}

\newpage
\baselineskip=24pt
\setcounter{page}{1}
\renewcommand{\thefootnote}{\arabic{footnote}}
\addtocounter{footnote}{-3}

It has long been known that symmetry may be restored at high
temperature in local thermodynamic equilibrium (LTE) \cite{ldjw}.
Recently it was realized that certain nonequilibrium (NEQ) conditions
can be even more efficient for symmetry restoration \cite{sr}.  An
example of such a nonequilibrium state can arise naturally after
inflation in the so-called preheating era \cite{KLS,pr}.  In fact,
symmetry may be restored in the NEQ state even if it is not restored
in the LTE state formed by thermalization of the NEQ state.  Detailed
numerical studies \cite{numr} confirm that 
fluctuations of inflaton decay products is large enough for
symmetry restoration, as well as for several other important effects,
including baryogenesis \cite{klr}, and supersymmetry breaking
\cite{alr}, and generation of a background of relic gravitational waves \cite{ktgw}. 

States with properties similar to those in preheating, namely,
anomalously large fluctuations and highly NEQ conditions, can arise in
other situations as well.  It was suggested in Ref.\ \cite{kr96} that
if bubble collisions produce large numbers of soft scalar particles
carrying quantum numbers associated with a spontaneously broken
symmetry, the phenomenon of (or tendency toward) symmetry restoration
may occur.  The basic point is that bubble collisions create NEQ
conditions with a large number of ``soft'' quanta of average energy
smaller than the equivalent LTE temperature corresponding to
instantaneous conversion of the bubble energy density into radiation.
Since it may require several scattering times for the low-energy
quanta to form a thermal distribution, it is rather reasonable to
consider the NEQ period as a separate epoch.  This is generally
referred to as the `preheating' epoch in a manner similar to the
preheating phase of slow-roll inflation \cite{KLS}.

The tendency of symmetry restoration in NEQ conditions after bubble
collisions may be readily understood by the following (somewhat
oversimplified) reasoning.  Let us imagine that particles $\chi$ are
produced in the bubble wall collisions and are charged under some
symmetry group, so that their mass, $m_\chi$, depends upon some scalar
field $\phi$ (the order parameter of the symmetry) as $m_\chi^2(\phi)
= m_0^2+g\phi^2$.\footnote{Of course $\chi$ particles may coincide
with the $\phi$ particles themselves, but in this example the
colliding bubbles are {\bf not} made from the field $\phi$. Otherwise
there can be some effect, but the original symmetry will not be restored.}
Here, $g$ represents a combination of numerical factors and a coupling
constant.  As a simple example we might assume that the
$\phi$-dependent mass originates from a potential term of the form
$V_{\chi\phi} = (1/2)g\phi^2\chi^2$.  As opposed to the large-angle
scattering processes required for thermalization, forward-scattering
processes do not alter the distribution function of the particles, but
simply modify the dispersion relation. This is true in NEQ conditions,
as well as the familiar LTE conditions. Forward scattering is
manifest, for example, as ensemble and scalar background corrections
to the particle masses.  Since the forward-scattering rate is usually
larger than the large-angle scattering rate responsible for
establishing a thermal distribution, the nonequilibrium ensemble and
scalar background corrections are present before the initial
distribution function relaxes to its thermal value.  These
considerations allow us to impose the dispersion relation
$\omega^2={\bf p}^2+m_\chi^2(\phi)$ for NEQ conditions.  

The leading contribution of the particles created by bubble collisions to the
one-loop effective potential of the scalar field $\phi$ can be shown
to be $\Delta V(\phi)\simeq (n/E)m_\chi^2(\phi)$ \cite{sr,rt96}, where
$n$ and $E$ are the number density and the energy of the $\chi$
quanta, respectively. We may write the potential for the NEQ
configuration as $\Delta V(\phi) = B_{{\rm NEQ}}\phi^2$, where
$B_{{\rm NEQ}}=gn/E$. In NEQ conditions, the coefficient $B_{{\rm
NEQ}}$ may be quite large, indeed larger that the corresponding
equilibrium coefficient which scales like $T_{{\rm
RH}}^2$, $T_{{\rm RH}}$ defined as the temperature of the universe
when the thermal spectrum of radiation is first obtained. Therefore,
the tendency of symmetry restoration may turn out to be rather
independent of $T_{{\rm RH}}$. We also notice that since the energy
$E$ scales like the inverse of the bubble wall width $\Delta$, $E\sim
\Delta^{-1}$, one can suggest that the effect of soft particles on
symmetry restoration is stronger for thick bubble walls.

The aim of the present paper is to investigate numerically the issue
of symmetry restoration in bubble wall collision.  We will explicitly
show that at the final stage of first-order phase transitions when
bubble collisions occur, nonthermal quanta are produced, and that they
tend to restore symmetry. This tendency can be quantified as a
shift of the order parameter $\phi$ from its equilibrium value
toward smaller values.   We
will also confirm the conjecture about the dependence of the strength
of symmetry restoration upon the bubble wall width. Finally, we will
comment on the possible implications that our result may have for
electroweak baryogenesis.

Let us concentrate on a theory with a single scalar field $\phi$ (the
$\chi$ particles of the above discussions must be identified with the
$\phi$) with Lagrangian density
\begin{equation}
{\cal L} =\frac{1}{2}\partial \phi_\mu \partial \phi^\mu
 -\frac{1}{2}m^2 \phi^2 + \frac{1}{3} c \phi^3 -
\frac{1}{4} \lambda \phi^4- V_0,
\label{e1}
\end{equation}
where $V_0$ is a constant.  We introduce the dimensionless variables
$\varphi \equiv \phi/ \phi_0 $, $ \tau \equiv \sqrt{\lambda} \phi_0 \,
t $, and $\xi = \sqrt{\lambda} \phi_0 \, x$, where $\phi_0$ will be
fixed later. In the new variables the factor $\lambda \phi_0^4$ is an
overall multiplication factor for the Lagrangian
($\widetilde{m}=m/\sqrt{\lambda}\phi_0$, $\widetilde{c} =
c/\lambda\phi_0$, $\widetilde{V_0} = V_0/\lambda\phi_0^4$)
\begin{equation} 
{\cal L} =\lambda \phi_0^4 
\left[\frac{1}{2}\partial \varphi_\mu \partial \varphi^\mu
 -\frac{1}{2}\widetilde{m}^2 \varphi^2 + \frac{1}{3} \widetilde{c} 
\varphi^3 -\frac{1}{4} \varphi^4  - \widetilde{V_0}\right]\equiv 
\lambda \phi_0^4 \left[\frac{1}{2}\partial \varphi_\mu \partial \varphi^\mu -
V(\varphi) \right] \ .
\label{e2}
\end{equation}
The overall factor will not enter the equation of motion.  The final
step is a choice of a potential, which we choose such that
$dV/d\varphi =\varphi\, (\varphi -1)\, (\varphi- \varphi_m)$. The
equation of motion is then
\begin{equation}
\label{e3}
\Box \varphi + \varphi\, (\varphi -1)\, (\varphi- \varphi_m)  =  0 \ .
\end{equation}
With this choice of $dV/d\phi$ the extrema of the potential
are transparent: it has minima at $\varphi=0$ and $\varphi=1$ and a
local maximum at $\varphi = \varphi_m$ (we thus fix the parameter
$\varphi_m$ to be in the range $0 < \varphi_m < 1$). We shall assume
$\varphi =1$ corresponds to the true vacuum, i.e., $V(0) >
V(1)$. Making the connection with Eq.\ (\ref{e1}), we conclude that
$\phi=\phi_0$ is the field strength in the true vacuum, and the
constants entering Eq.\ (\ref{e1}) are $m^2 = \varphi_m\, \lambda
\phi_0^2$ and $c = (1 + \varphi_m)\, \lambda \phi_0$.  We shall
require the absence of cosmological constant in the true vacuum, $V(1)
=0$; this gives $V_0 = (1-2 \varphi_m)\lambda \phi_0^4/12$. Since we
consider the true vacuum to be at $\varphi=1$ and the false vacuum at
$\varphi=0$, we can further restrict the parameter $\varphi_m$ to be
in the range $0 < \varphi_m < 0.5 $.

Note that only one parameter, $\varphi_m$, enters the equation of
motion in the rescaled variables. This is a key point.  The evolution
of any initial field configuration, $\varphi (\tau=0,\xi)$, for fixed
$\varphi_m$ will be the same in the rescaled variables, regardless of
the coupling constant $\lambda$.

The initial field configuration for the problem at hand corresponds to
a set of new-phase critical bubbles expanding in the false vacuum. Note
that the evolution of a critical bubble is also defined by Eq.\
(\ref{e3}), and consequently it is fixed when $\varphi_m$ is fixed.
However, the bubble nucleation probability is a more complicated
function of the other variables (note that nucleation became
unsuppressed when $\varphi_m \rightarrow 0$, i.e, when the potential
barrier disappears). The nucleation probability will determine the
initial separation of critical bubbles (in space, as well as in
time). In our numerical integration we will consider the mean
separation of bubble nucleation sites as another free parameter of the
model . Fixing it gives one extra constraint on the set of parameters
$\lambda$, $\phi_0$ and $\varphi_m$.

After nucleation, new phase bubbles expand and collide.  After
collisions, the spatial distribution of the magnitude of $\varphi$
resembles a random superposition of many wavelength modes---a
configuration with large field fluctuations.  It is important that the
system is classical and can be described by Eq.\ (\ref{e3}) from the
time of bubble nucleation, through the time of bubble collisions and
the condition of large field fluctuations.  

The random-wave configuration is quickly established after bubble
collisions; essentially it is established on the time-scale of bubble
collisions since there is no small parameters in Eq.\
(\ref{e3}). Eventually the waves interact and LTE is established.  Since
transforming the NEQ distribution function into an LTE distribution
function involves producing states with small occupation number, the
coupling constant $\lambda$ will enter the time scale for the
establishment of LTE. This time scale can be very long if $\lambda$ is
small, so the NEQ configuration can exist for a long time. This phase
has specific properties which are the subject of our study here.

First, let us recall what is expected in the final LTE state.  The LTE
temperature can be found using energy conservation
\begin{equation}
g_*\frac{\pi^2}{30}T^4_{\rm LTE}=V_0 = \lambda \phi_0^4 
\frac{(1-2 \varphi_m)}{12} \ ,  
\label{e5}
\end{equation}
which gives 
\begin{equation}
T_{\rm LTE}=\left(\frac{\lambda}{g_*}\right)^{1/4} \phi_0 \left[ 
\frac{5(1-2 \varphi_m)}{\pi^2}\right]^{1/4}
\equiv \lambda^{1/4} \phi_0 \, b,
\label{e6}
\end{equation}
where $b$ is a constant of order unity and $g_*(T)$ is the number of
relativistic degrees of freedom at temperature $T$.  Note that $T_{\rm LTE}$
approaches zero as $\lambda$ approaches $0$. Due to interactions with
the medium, LTE values of the model parameters, e.g., the effective mass, are
different than vacuum values.  The value of the parameters can be
calculated as loop corrections to the action. Most important is the
change of the effective mass, $m_{\rm eff}^2 (T) = m^2 + \lambda T^2/4
$. At very high temperatures $m_{\rm eff}^2 (T)$ becomes positive,
even if the zero-temperature value of $m^2$ was negative. This is a
signal that broken symmetries are restored at high temperatures
\cite{ldjw}.

In the model of Eq.\ (\ref{e1}) which we consider here, the symmetry
can not be restored again after bubble collisions, but the temperature
dependent contribution to the effective mass will be nonzero. Using
Eq.\ (\ref{e6}) we find that it scales with coupling constant as
$\lambda^{3/2}$, and tends to zero as $\lambda$ tends to $0$.  Note
for what follows that the temperature-dependent correction to the mass
can be written in more general form, as $m_{\rm eff}^2 = m^2 + 3
\lambda \langle \phi^2 -\langle \phi \rangle^2 \rangle $.

Let us now find the mean value of the field $\varphi$ in thermal
equilibrium with temperature given by Eq.\ (\ref{e6}).  To leading
order in the coupling constants, the equation $dV_{\rm eff}/d\varphi
=0 $ becomes
\begin{equation}
(\varphi_m + 3\sqrt{\lambda} b^2) \varphi -(1+ \varphi_m ) \varphi^2
+ \varphi^3 -(1+\varphi_m) \sqrt{\lambda} b^2 /12 = 0  \,  ,
\label{e7}
\end{equation}
where terms proportional to $\sqrt{\lambda}$ are reminiscent of
temperature-dependent corrections to the effective potential rewritten
in terms of our dimensionless variables. We see that the solution of
this equation tends to $\varphi =1$ when $\lambda \rightarrow 0$. In
other words, the mean value of the field $\varphi$ in thermal
equilibrium (established after the phase transition is completed)
differs very little from the vacuum expectation value if the coupling
constant is small.

We can study the process of bubble collisions and subsequent
chaotization by numerically integrating Eq.\ (\ref{e3}). We defined a
3-dimensional box of size $l$ on a grid of size $128^3$ employing
periodic boundary conditions.  With periodic boundary conditions every
bubble in the box is mirrored by its (infinitely repeating)
reflections.  As the bubble expands to fill the box, it will collide
with its reflections, and there is no need to put more than one bubble
inside the box to study bubble collisions. So we have restricted
ourselves to an initial configuration corresponding to just one
critical bubble of the true phase in the box.  In this case the size
of the box, $l$, corresponds to the mean initial separation of bubbles
in units of $1/\lambda^{1/2}\phi_0$. We integrated the equation of
motion for $l=4,8,10$ and $12$, corresponding to progressively larger
bubbles at collision time.  We used $\varphi_m =0.1$ for the only
parameter in the equation of motion.

\begin{figure}
\psfig{file=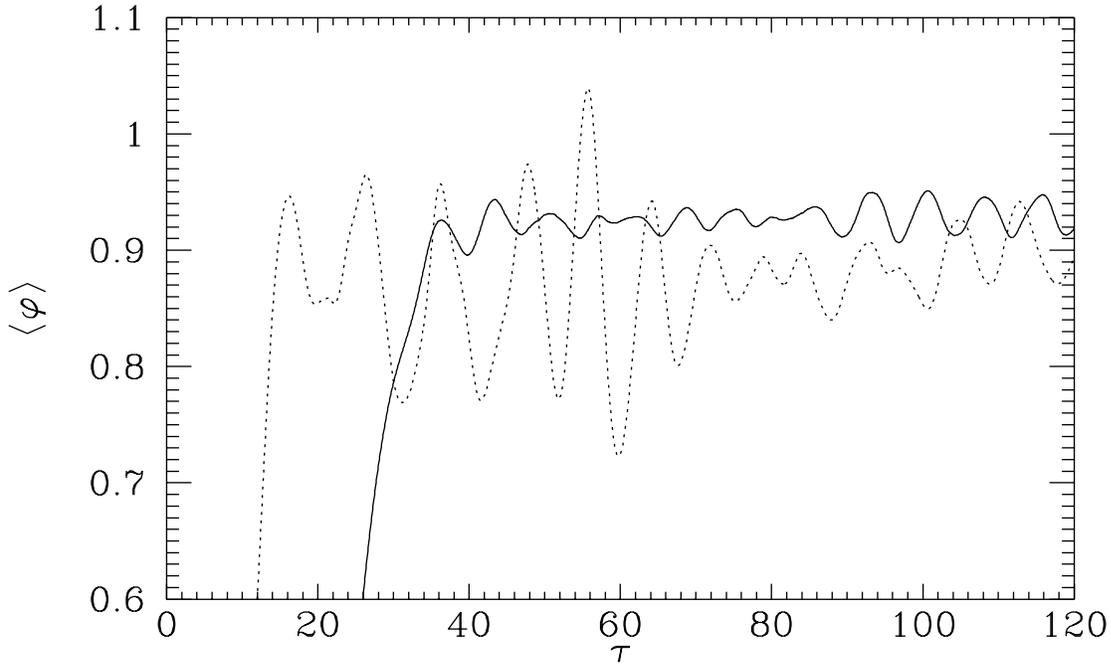,height=3.8in,width=6in}
\caption{Time dependence of the zero-momentum mode, $\varphi_0 =
\langle \varphi \rangle$. The dotted line corresponds to initial 
bubble separation
of $l =4$, the solid line corresponds to $l =8$.}
\label{fig:zmod}
\end{figure}

The results for the time dependence of zero mode of the field,
$\varphi_0 = \langle \varphi \rangle$, is presented in Fig.\
\ref{fig:zmod}, where $\langle \dots \rangle$ means the spatial
average (over grid points).  We see that after bubbles have collided
($\tau> 16$ for $l = 4$ and $\tau > 40$ for $l =8$), the zero mode
does not relax to is vacuum value, $\varphi_0 =1$, but oscillates near
some smaller value. We define $\varphi_0 \equiv \overline{\langle
\varphi \rangle}$, where bar denotes the time average over several
oscillations. We find $\varphi_0 \approx 0.93$ in the case $l=8$ and
$\varphi_0 \approx 0.87$ with $l=4$ at $\tau \sim 80$. Note that
$\varphi_0$ rises slightly with $\tau$, which is the sign of ongoing
relaxation. We do not present results for $l=10$ and $l=12$ since they
do not differ appreciably from the case $l=8$ ($\varphi_0$ at $l=12$
is larger by an about $0.01$ than the corresponding value for $l=8$).

\begin{figure}
\psfig{file=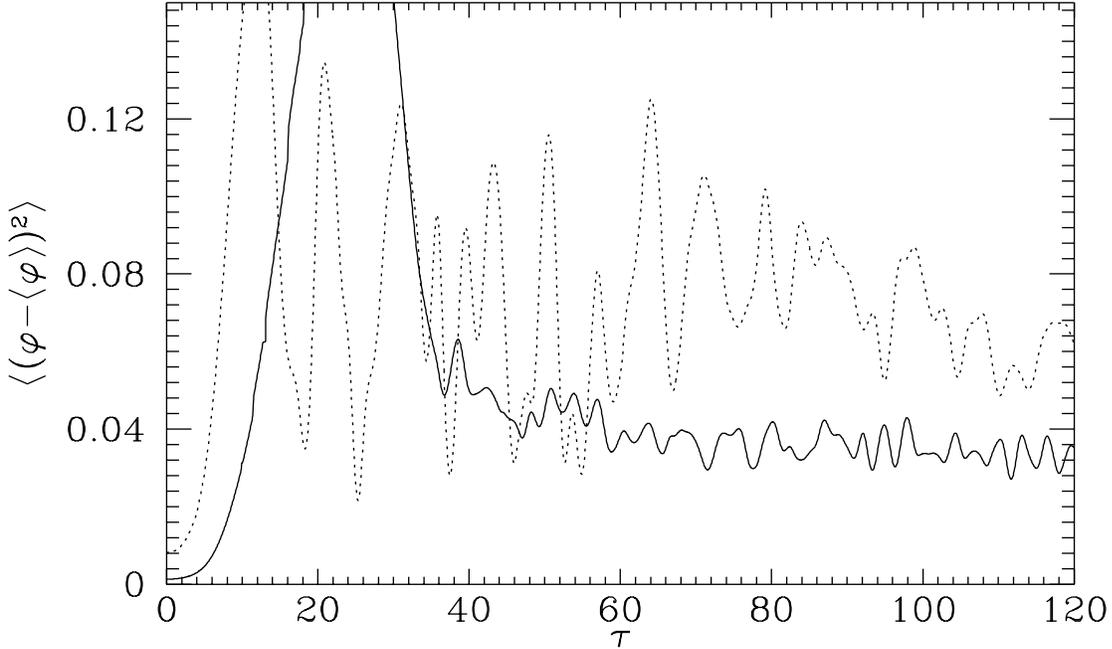,height=3.8in,width=6in}
\caption{Time dependence of the variance, $\langle \varphi^2 \rangle  -
\langle \varphi \rangle ^2$. The dotted line corresponds to initial  bubble 
separation of $l =4$, the solid line corresponds to $l =8$..}
\label{fig:var}
\end{figure}

The deviation of $\varphi_0$ from the vacuum value is not unexpected
since a random field of classical waves is created after bubble
collision, i.e., ${\rm Var} (\varphi) \equiv \langle \varphi^2 \rangle
-\langle \varphi \rangle ^2$, is nonzero. The time dependence of the
variance is shown in the Fig.\ \ref{fig:var}. Note again that with
fixed initial conditions the variance does not depend upon $\lambda$,
i.e., it has a nonthermal origin.

At $\tau \sim 80$, with $l=8$ we have ${\rm Var} (\varphi) \approx
0.036$ and with $l=4$ we find ${\rm Var} (\varphi) \approx 0.08$.
Again we employ time averaging over several oscillations. At small
$\lambda$ those values are much larger than its equivalent LTE value
${\rm Var} (\varphi) = T^2_{\rm LTE} /12$ (see Eq.\ (\ref{e6})). The
fact that ${\rm Var} (\varphi)$ in NEQ can exceed its equivalent LTE
value by many orders of magnitude was the main point of Ref.\
\cite{sr} which studied the preheating phase after inflation, and of
Ref.\ \cite{kr96} which studied conditions following bubble collisions.
Our work supports the claim in Ref.\ \cite{kr96} that NEQ phase
transitions can occur in models which contain more degrees of freedom 
than the simple toy model of Eq.\ (\ref{e1}).

Let us see whether we can understand the deviation of the zero mode
from its vacuum value by the existence of a nonzero ${\rm Var}
(\varphi )$. Let us decompose the field as $\varphi = \varphi_0 +
\delta \varphi$, and substitute this decomposition into the equation
$dV/d\varphi =0$.  We find in the Hartree approximation
\begin{equation}
(\varphi_m + 3\langle \delta\varphi^2 \rangle) \varphi_0 -(1+ \varphi_m ) 
\varphi_0^2 + \varphi_0^3 
-(1+\varphi_m)\langle \delta\varphi^2 \rangle = 0 \, . 
\label{e8}
\end{equation}
Assuming in addition the deviation of $\varphi_0$ from 1 to be small,
we find
\begin{equation}
\varphi_0 =1 -\frac{2-\varphi_m}{1 - \varphi_m + 3 \langle \delta \varphi^2 \rangle} 
\langle \delta\varphi^2 \rangle   \, . 
\label{e9}
\end{equation}
Using $\varphi_m =0.1$ and the values of $\langle \delta\varphi^2
\rangle$ inferred from Fig.\ \ref{fig:var}, we find $\varphi_0 =0.93$
for $l=8$ and $\varphi_0 =0.87$ for $l=4$, which are in excellent
agreement with the results presented in Fig.\ \ref{fig:zmod}.

We can also understand the dependence upon $l$, the initial bubble separation.  
The larger the initial bubble separation, the longer bubbles will expand
before they collide.  As a bubble expands, its wall thickness decreases.  Hence, 
colliding bubbles in the $l=8$ calculation are thinner than in the $l=4$ case.  Following the
discussion in Ref.\ \cite{kr96}, we expect the average energy of the quanta created
in wall collisions to scale as $\Delta^{-1}$, where $\Delta$ is the wall thickness at collision.
Since the effect of the background on the effective potential scales as 
$n/ E \propto \Delta$, we expect the $l=4$ calculation to result in a larger departure
from the vacuum value.  This expectation
is confirmed by the results shown in the figures.
  
Even though we only examine a particularly simple model, we conjecture
that a deviation of $\varphi_0$ from its thermal equilibrium value in
the aftermath of bubble collisions may have important consequences for
some applications of first-order phase transitions, e.g., electroweak
baryogenesis. In any scenario where the baryon asymmetry is generated
during a first-order electroweak phase transition, the asymmetry is
generated in the vicinity of bubble walls, and a strong constraint on
the ratio between the vacuum expectation value of the Higgs field
inside the bubble and the temperature must be imposed,
$\langle\phi(T)\rangle/T>1$ \cite{ew1}. This bound is necessary for
the just created baryon asymmetry to survive the anomalous baryon
violating interactions inside the bubble, and may be
translated into an severe upper bound on the physical mass of the
scalar Higgs particle. Combining this bound with the LEP
constraint already rules out the possibility of electroweak
baryogenesis in the standard model of electroweak interactions,
and even leaves little room for electroweak baryogenesis in
the minimal supersymmetric extension of the standard model
\cite{ew5}. Since the rate of anomalous baryon number violating
processes scales like ${\rm exp}(-\langle\phi\rangle /T)$, it is clear
that even a small change in the vacuum expectation value of the Higgs
scalar field from its equilibrium value may be crucial for electroweak
baryogenesis considerations.  Our results suggest that imposing the
bound $\langle\phi(T)_{{\rm EQ}}\rangle/T>1$ may not be a sufficient
condition for successful electroweak baryogenesis. NEQ effects at the
completion of the phase transition may reduce the expectation
value of the Higgs field, thus enhancing the anomalous baryon number
violating rate with respect to its equilibrium value, making the upper
bound on the Higgs mass more severe. Applications of our
considerations to the electroweak transition may result in a fatal
blow to many scenarios involving extensions of the standard model
where the baryon asymmetry is generated during the electroweak phase transitions.

The model we consider in this paper is quite simple, but it illustrates several points.  
The most important result is that NEQ conditions following bubble collisions
can have a dramatic effect upon the effective potential.  Although the model we
study is too simple to result in symmetry restoration, the numerical results confirms
the assumptions made in Ref.\ \cite{kr96} about the efficiency of NEQ conditions.
We mentioned a possible direct application of our results to electroweak baryogenesis,
but we believe that the phenomenon of NEQ effects will have other implications as well.

\vspace{24pt}

The work of EWK and AR was supported in part by the Department of
Energy and by NASA grant NAG 5-2788.  IT was supported by Department
of Energy grant DE-AC02-76ER01545 at Ohio State.

\end{document}